\documentclass[runningheads]{llncs}

\usepackage[T1]{fontenc}
\usepackage{graphicx}
\usepackage{cite}

\begin{document}

\title{Allies of Atargatis: A VR Game for Healing the Eastern Mediterranean Sea and the Self}

\titlerunning{Allies of Atargatis}

\author{Casondra Sobieralski\inst{1}\orcidID{0000-0002-7022-4216} \and
Hector Fan\inst{2}\orcidID{0009-0009-4303-3818}}

\authorrunning{Sobieralski and Fan.}

\institute{Northeastern University, Oakland, CA 94613, USA 
\email{c.sobieralski@northeastern.edu}
\url{https://casondra.pelefire.com/}
\and
Northeastern University, Boston, MA 02115, USA
\email{fan.huaf@northeastern.edu}\\
\url{https://www.hectorfan.com/}}



\maketitle

\begin{abstract}
``Allies of Atargatis'' is a meditative, posthumanist Virtual Reality game about working collaboratively across borders to heal the Eastern Mediterranean Sea from the effects of adverse environmental impacts. The game suggests a relationship between self care and healing the sea. The breath-driven interface design, soothing environment, and game mechanics facilitate deep relaxation, a sense of safety, and mind-body awareness. This game is for anyone interested in marine ecology or mind-body wellness, but the main audiences motivating this project are: pre-teens in the Eastern Mediterranean region suffering from anxiety born of war, terrorism, and refugee displacement; youth visitors to Eastern Mediterranean aquariums; and–in consultation with a pediatric trauma surgeon at University of Pittsburgh Medical Center–youth who need soothing before and after surgery. The game concept is inspired by the work of the Arava Institute for Environmental Studies, which brings Palestinian, Israeli, and Jordanian students together to solve environmental problems.



\end{abstract}

\section{Introduction: About Allies of Atargatis, Intended Audiences}

\begin{figure}[t]
    \centering
    \includegraphics[width=\textwidth]{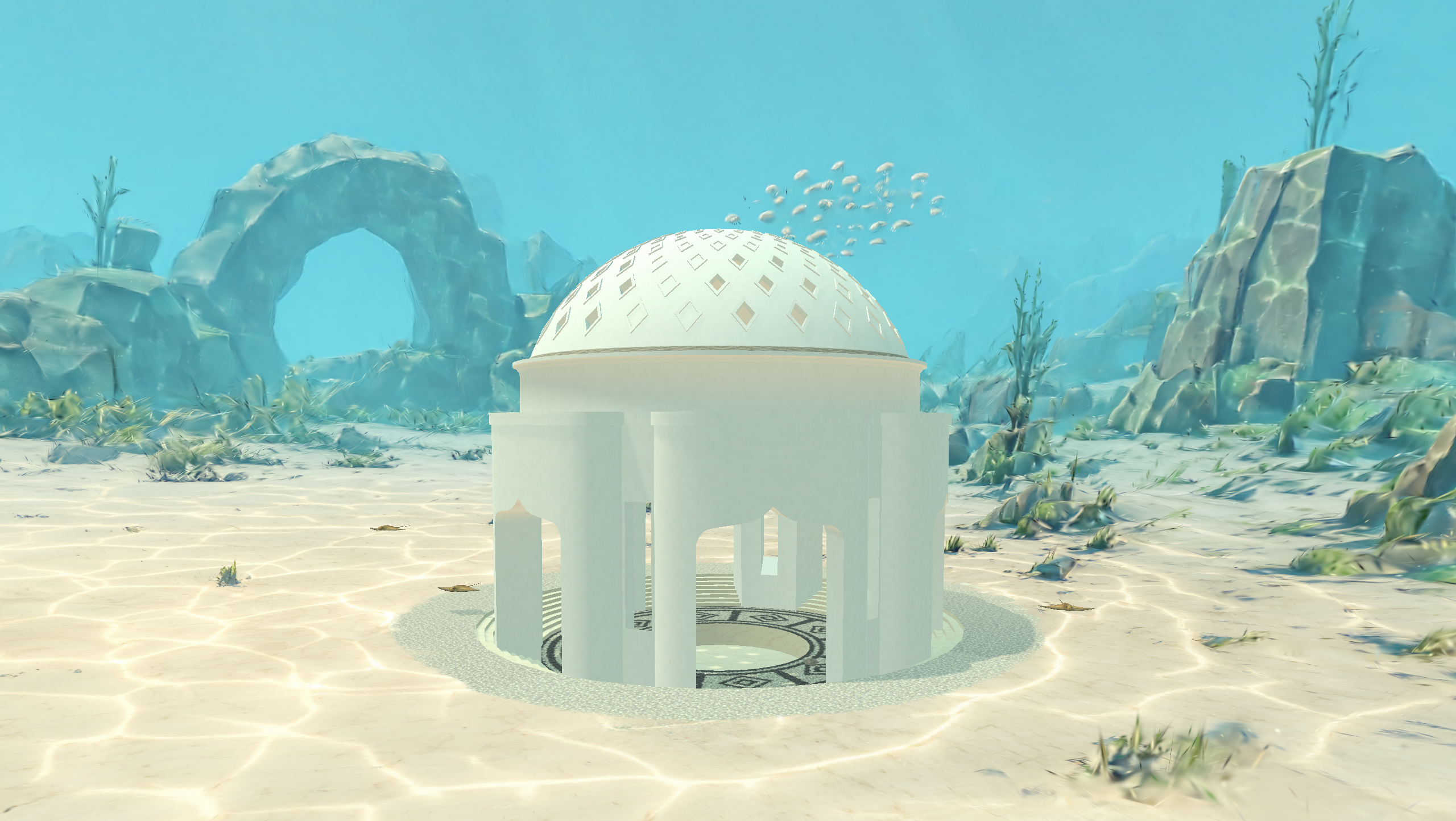}
    \caption{A rendered 3D model of the underwater sea spa, based on an architectural structure in the hydrotherapy spa town of Kallithea on the island of Rhodes, Greece.}
    \label{fig:teaser}
    \
\end{figure}

``Allies of Atargatis'' is a meditative, posthumanist Virtual Reality game in which players work to heal the Eastern Mediterranean Sea from the effects of adverse environmental impacts while also healing themselves. The breath-driven interface design and soothing environment facilitate deep relaxation, a sense of safety, and interoceptive (mind-body) awareness. Game mechanics reward slowing down, empathy, cooperation with sea creatures, and mutual nurturing. 

As the Eastern Mediterranean waters warm, species such as lionfish are relocating from the Red Sea via the Suez Canal. Some of these non-native species proliferate madly and voraciously devour everything in sight, thereby wreaking havoc on Eastern Mediterranean marine ecosystems. Plastics are killing marine life. Specific sound frequencies are destroying dolphins’ lungs. Turtles get caught in fishing nets, leaving jellyfish populations unchecked.\footnote{Personal interviews with marine scientists Vivian Louizidou and Gerasimos Kondylatos at the Hydrobiological Station of Rhodes, June 3 and June 11, 2024.} Players work with virtual sea creatures of the region to take agency in solving these problems and to heal the marine environment.

A mermaid goddess, Atargatis, guides players on this quest. Atargatis was a Bronze Age Semitic fish-tailed deity worshiped in the region known in antiquity as Syro-Palestine. She represents love, the sea, ``feminine powers,'' protection of both cities and people…and mourning. Archaeologists have found evidence of her veneration in Ashkelon (today part of Israel), Hierapolis (today part of Syria), and north of Petra (in today’s Jordan). The popularity of mermaids eventually spread through the Mediterranean.

Here and now, the mermaid’s urgent message to players is, ``We saw rising sea levels and climate crises in the Eastern Mediterranean during the Late Bronze Age, and it did not bode well for civilizations.'' Archaeologists theorize that environmental factors–perhaps a solar maximum that resulted in a mega-drought–provoked varying degrees of system collapse, mass migrations, and resource wars\cite{Cline_CollaCultu_2024}. With our current climate change crisis, scientific models forecast that sea levels around the Eastern Mediterranean could rise .43 to 2.5 meters by 2100 CE as regional waters warm 20 percent faster than the global average\cite{UNEnviron_ClimaChang_2021}.

Hence, Atargatis encourages us to learn from the past. The mermaid also reminds us that there is a mutual, overarching threat to all of the coastal modern geopolitical entities that once comprised Syro-Palestine–Syria, Lebanon, Palestine, Israel–as well as to Greece, Turkey, and Cyprus. The mermaids offer a feminist alternative to warring with one another, which is pulling together to heal and protect that which sustains all life: the sea. The sea connects us.

\begin{figure}[t]
    \centering
    \includegraphics[width=\textwidth]{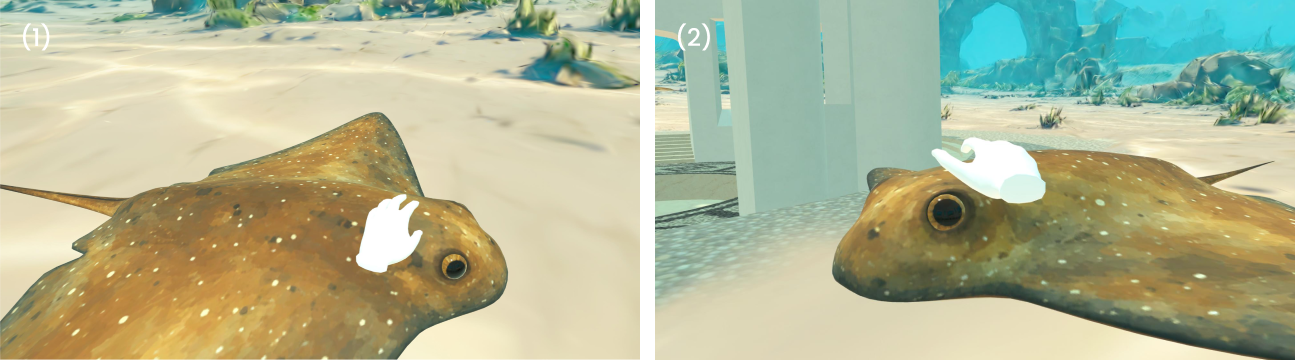}
    \caption{Renderings from within the VR world showing the player petting common eagle rays as they swim around the underwater sea spa.}
    \label{fig:ray}
\end{figure}

While working to heal the sea, players will also be healing themselves through regulating breath, floating within soothing imagery, and enjoying soundscapes. Game mechanics are designed such that players gain rewards for maintaining a languid pace that mirrors the behavior of indigenous fish species of the region; this is counter to the demeanor of some aggressive non-native fish. Players also gain rewards for practicing empathy and mutual exchanges of care with non-human helpers at an underwater sea spa. When players slow down to pet eagle rays, the rays benefit and the player gains life force points (Fig.~\ref{fig:ray}).

``Allies of Atargatis'' is for anyone interested in marine ecology, healing the environment, mind-body wellness, and/or global citizenship. However, the main audiences motivating this project are: 1) pre-teens in the Eastern Mediterranean region–insofar as electricity is available–suffering from anxiety born of war, terrorism, and refugee displacement, 2) youth visitors to Rhodes aquarium, and 3) youth who need soothing before and after surgery. The latter demographic is part of a project with a pediatric surgeon at the University of Pittsburgh Medical Center. ``Allies of Atargatis'' is inspired by the work of the Arava Institute for Environmental Studies, which brings Palestinian, Israeli, and Jordanian students together to solve environmental problems. Their moto is, ``Nature knows no political borders\cite{AravaInst_AravaInsti_2026}.'' 

\section{The Genesis of Allies of Atargatis}
This project evolved from the Director of ``Allies of Atargatis'' processing her own grief over current cycles violence in the Levant that have crescendoed since October 7, 2023.  History demonstrates that trauma begets violence begets trauma begets violence in the Middle East. Can a healing game offer even the tiniest feminist intervention in this cycle that has lasted for over a century?

At the start of the war, the Director kept doodling mermaids when she felt waves of fear. She started to wonder, where does the myth of the mermaid come from? Cursory research revealed that mermaids seem to have originated in the Levantine region as representations of the fish-tailed goddess Atargatis. Archaeologists, however, are not entirely sure whether Atargatis is a distinct goddess or, rather, a facet of the Canaanite/Phoenician goddess Astarte. Astarte was akin to the Mesopotamian Ishtar–a goddess of war and battles, but also of love and unbridled passion, universal power, the source of all life. She was associated with the sea and mountains\cite{Karageorgh_KypriAphro_2005}. Perhaps not coincidentally, the Director heavily researched Astarte for her 2022 PhD dissertation project. That three-year endeavor resulted in a meditative archaeology VR game. Thus, ``Allies'' builds on knowledge gained from that prior project.



\section{Process of Researching Game Content}

\begin{figure}[htbp]
  \centering
  \includegraphics[width=1\linewidth]{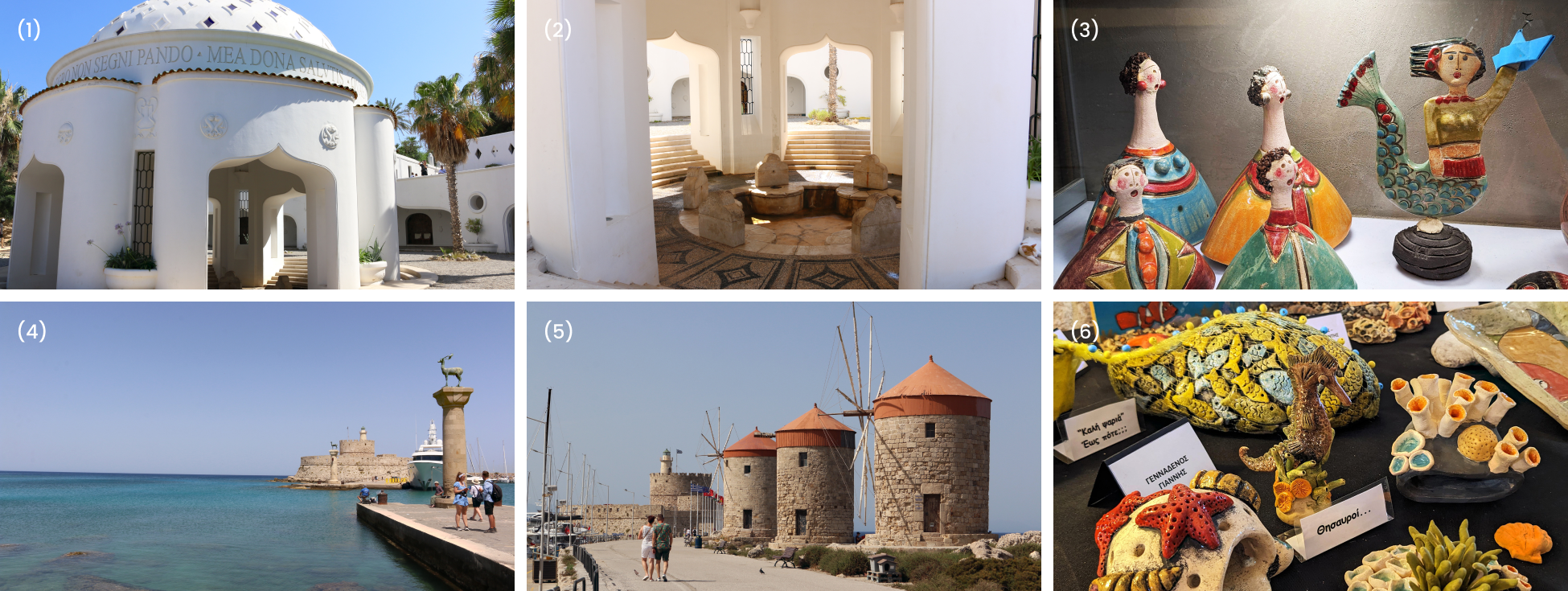}
  \caption{Media assets captured on the Greek island of Rhodes inform worldbuilding for ``Allies of Atargatis.'' Represented here are: Kallithea architecture for the underwater sea spa, the built environment of Rhodes Harbor, and locally made art to inspire stylization within the game.}
  \label{fig:fieldwork}
\end{figure}

During the summer of 2024, the Director interviewed marine scientists at the Rhodes Hydrobiological Station, researched native and non-native Eastern Mediterranean species at the affiliated Rhodes Aquarium, captured underwater audio and video off the coast of Rhodes. She also captured media assets documenting the architecture of Rhodes–a pastiche of Classical, Byzantine, Ottoman, and Modernist forms–to inform worldbuilding. A small, temple-like structure in the hydrotherapy spa town of Kallithea (Fig.~\ref{fig:fieldwork}-1) informed the design of the VR game’s underwater sea spa  (Fig.~\ref{fig:teaser}); the gentle curves and perforated spaces of the structure beckon as gateways for fish, rays, and mermaids to swim through. Museums in Rhodes and Athens, including a museum of ancient toys and games, informed props to enhance worldbuilding in ``Allies.'' European websites and government reports on climate change and climate initiatives added to a baseline understanding of the scope of environmental damage occurring in the Mediterranean Sea. Marine life image databases and online videos from the BBC, Monterey Bay Aquarium, and scuba divers also informed 3D animation.

The posthumanist elements of the game drew from research that was part of the Director’s dissertation project, but also from recent works of fiction and creative non-fiction \cite{Imbler_HowFar_2022,Bjornerud_TurniStone_2024,Khatwa_WhispRock_2025,Schlanger_LightEater_2024,Montgomery_SoulOctop_2015}. How does one tell a story about science that engages the emotions and a sense of connection to non-human life? How do some scientists, as creative authors, embrace subjectivity and break human-nonhuman hierarchies?


\section{Design Strategies for Immersion, Posthumanist Interaction Experiences, and Education}

\subsection {Immersion}

Primary design goals for ``Allies of Atargatis'' include creating a convincingly immersive world, creating a relaxing sense of embodied presence, and breaking down hierarchies among human and non-human characters to establish a sense of mutuality and a posthumanist consciousness. Additionally, the game seeks to educate players about problems and solutions regarding environmental threats to the Eastern Mediterranean Sea. 

\begin{figure}[htbp]
  \centering
  \includegraphics[width=1\linewidth]{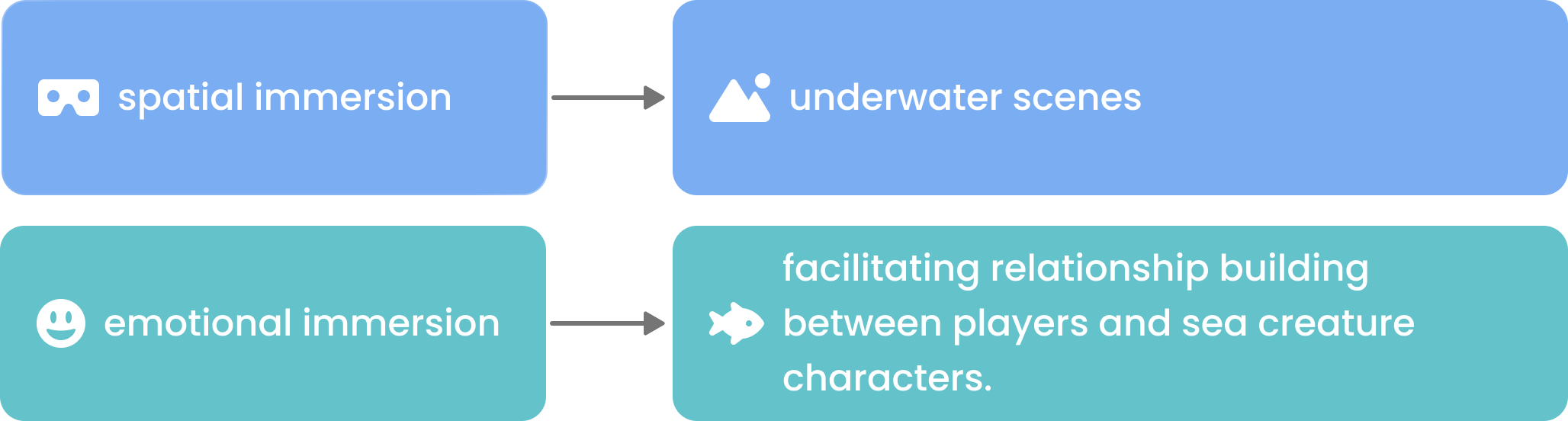}
  \caption{Diagram of primary types of immersion in ``Allies of Atargatis.''}
  \label{fig:immersion}
\end{figure}

\begin{figure}[htbp]
  \centering
  \includegraphics[width=1\linewidth]{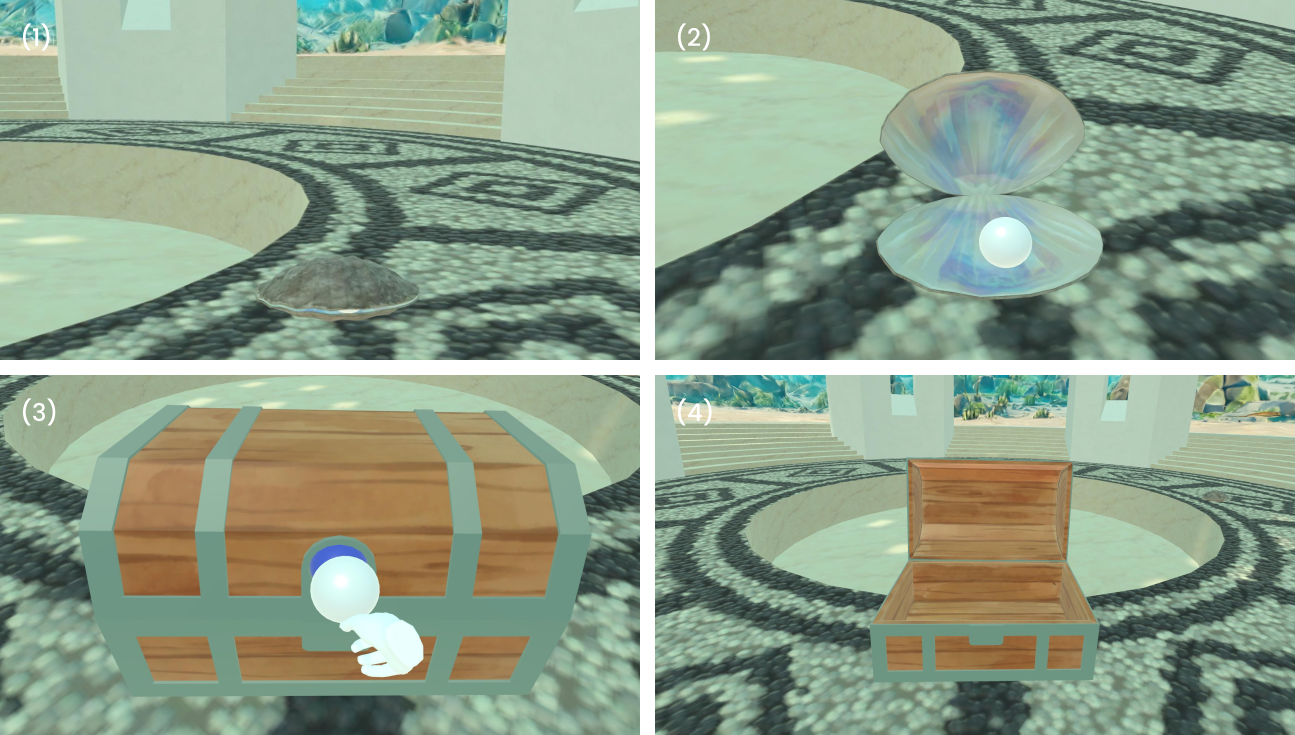}
  \caption{Renderings from within the VR world: The player reaches for pearls within oysters. Pearls are used like currency to unlock treasure chests of knowledge and, eventually in Phase 2, ancient ``vending machines'' of sedative fish food to tame lionfish.}
  \label{fig:pearl}
\end{figure}

In establishing means to create a sense of immersion, Marie-Laure Ryan and Janet Murray inform. Ryan categorizes types of immersion, noting that spatial immersion is a response to setting, temporal immersion is a response to story, and emotional immersion is a response to characters.\cite{Ryan_NarraVirtu_2015}. ``Allies'' is particularly focused on spatial immersion via underwater scenes and emotional immersion via facilitating relationship building with sea creature characters. Both game strategies encourage what Ryan defines as narratological metalepsis, a sense of becoming a character in the story one reads, watches, or enters\cite{Ryan_NarraVirtu_2015}. Ryan also sees ritual and manipulation of symbolic objects as means of establishing immersion\cite{Ryan_NarraVirtu_2015}. Players in ``Allies'' engage with symbolic objects such as pearls of wisdom that empower them to unlock knowledge from treasure chests (Fig.~\ref{fig:pearl}). Murray advises that a regulation of the boundary between the imaginary and the real is necessary for a sense of immersion to be maintained. How is a participant cued as to when they are entering and exiting the fantasy space, the so-called ``magic circle''\cite{Murray_HamleHolod_2017a}? In ``Allies,'' we designed an interstitial zone, above the water, at the Harbor of Rhodes. There the player transitions from ordinary ``reality'' to the game world as Atargatis explains the player’s mission and the reason for it. Once the player submerges, she enters a non-ordinary realm. However, she can resurface at the harbor at any point to avoid claustrophobic feelings. 

Other immersion strategists who have influenced this game focus on worldbuilding. Henry Jenkins urges game designers to consider a sense of place, to think of design as narrative architecture\cite{Jenkins_GameDesig_2004a}. John Bucher also stresses environmental storytelling, i.e. building a VR story from its narrative space\cite{Bucher_StoryVirtu_2018}. Mark J.P. Wolf suggests that symbolic signifiers of time, place, and culture, including artifacts, myths, rituals, and social maps, create a consistent, coherent world, e.g. cognitive immersion\cite{Wolf_BuildImagi_2012}. ``Allies'' obviously uses the Eastern Mediterranean Sea as its narrative space, and Atargatis herself speaks to the mythology of the world depicted. Architecture both above and beneath the water also indicate a culture, as do artifacts such as an ancient ``vending machine'' modeled after one in a Rhodes games and toys museum. In the VR space, the vending machine contains ``fish food,'' rather than the historically accurate content of wine, to pacify threatening lionfish. (The ``real world'' solution to lionfish overpopulation is introducing them on restaurant menus, but some children might find this upsetting.) 

Less concrete than artifacts and physical spaces in worldbuilding, but of utmost affective importance, is atmosphere. Underwater scenes have atmospheric power, and require deliberate atmospheric choices including materials, lighting, marine companions, and again artifacts/architecture that speak to time and place. We take cues from philosopher Gernot Böhme, who argues for an aesthetics that considers a relationship between environmental qualities and human states to create feeling and mood spatially\cite{Bohme_AesthAtmos_2017}. Yet, Böhme also stresses attention to material qualities that create ``atmospheric aura,'' convey a culture/time period, carry meaning and ideologies\cite{Bohme_AesthAtmos_2017}. To straddle these material and immaterial ways of being, he recommends: sound/acoustics, lifeforms, visual and spatial structures, and historical depth\cite{Bohme_AesthAtmos_2017}. In designing and modeling for ``Allies,'' we consider and employ all of these recommendations.    

Carrie Heeter, a game designer and yogini, advocates for creating a sense of immersion and presence within virtual worlds via interoception. Interoception involves a sensory awareness of bodily systems (like breath or heartbeat), the environment that the body occupies, and the interplay between the two. Heeter sees the potential of both meditative arts and Virtual Reality to enhance interoception\cite{Heeter_MeditMedit_2016}. Heeter’s ideas about embodied presence informed the design of a breath interface first for the Director’s earlier archaeology game, then we improved upon it for ``Allies.''\footnote{Thinking about use of breath for interface design is not totally unique, as artist Char Davies utilized a breath interface in her VR art piece ``Osmose'' in 1995\cite{Davies_Osmos_1995}, and the game ``Deep''\cite{Smit_DEEPVR_2021} uses a breath interface triggered by the  expansion and contraction of one’s rib cage (diaphragm) to help people overcome depression and anxiety. However, breath is still not a common mechanism for VR games, nor is the linkage of interface design with presence.} Combining Böhme’s ideas of atmospheres with Heeter’s ideas of interoception within an environment opens rich potentials for a soothing sea game, whereby water itself, sea life, and ancient signifiers all have sensuous material qualities.

\subsection {Posthumanist Interaction Experiences}

Another goal of ``Allies of Atargatis'' is to break down human/non-human binaries to encourage a posthumanist sense of interconnectedness with the ecology. The game seeks to suggest a paradigm of cooperation and empathy that challenges a paradigm of domination and hierarchy, the latter being the very paradigm that leads to cycles of both environmental degradation and wars. One way to instill an ethos of empathy and mutuality into the game mechanics for ``Allies'' was rejecting the idea of competitive gaming based on ``winning'' to instead embrace a strategy of cooperating – with sea creatures. 

The Director’s 2022 archaeology game was based on physicist Karen Barad’s ideas of quantum intra-action and agential realism, and ``Allies'' builds on this conceptual model. In simplified terms, intra-action recognizes interdependent webs of ever-changing (dynamic) relational events, or entanglements. This fluid model dismantles the typical hierarchical interaction design framework which assumes that independent entities (subjects) act upon other independent entities (objects)\cite{Barad_MeetiUnive_2007}. ``Allies'' instead gives cooperative, entangled agency to humans, mythical creatures, and marine life. The treasure chests of knowledge, unlocked with pearls, take on animate qualities and agency as they voice wise and knowing instructions for ecological missions within the marine environment. One could argue that these entangled agencies, not widely recognized in the dominant culture outside the diegetic world of the game, enhance immersion and are part of demarcating that the player is within the ``magic circle.'' 

Barad has influenced other interaction designers, such as Lucy Suchman. Suchman is an advocate of ``situated action.'' That is, she believes that interaction should be an embodied, adaptable, responsive process rather than a procedure\cite{Suchman_HumanRecon_2007}. Suchman’s ideas influence the idea of ``Allies'' as an exploratory game rather than a linear game: one explores the environment freely, and is encouraged to visit the sea spa as desired, or to increase life force points. Suchman encourages responsiveness to user/player needs\cite{Suchman_HumanRecon_2007}. This might be the need to surface if one feels claustrophobic underwater, the need to rest at the sea spa, or the desire to repeat an eco-task if a player has an affinity for working with a particular creature. The breath interface should also adapt to a player’s pace of breathing. In playtesting for just the breath interface element of her 2022 archaeology game, the Director found that there is no single rate of breathing that feels comfortable to all players. The interface should be flexible enough to make sure all players are comfortable. 

\subsection {Education and Knowledge Production: Aims of Serious Games}

Unlike a game intended solely for entertainment, a serious game needs to consider issues of knowledge production, subjectivity versus ``objectivity,'' and ``truthiness.'' The experiences and theories of cyber-archaeologists and digital heritage scholars/creators that informed the Director’s 2022 game can contribute considerations and solutions to any serious game. 
 
For example, Alice Watterson encourages the role of imagination, creativity, and storytelling in virtual archeology, here applied to ecology. She believes there is power in negotiating a complimentary partnership between subjective and objective methods/perspectives. However, in order to communicate or foster knowledge, audiences need to understand nuances of interpretation, storytelling, and display. I.e., how does the public know where subjective, creative, editorial processes come into play? People’s tendency is to cling to what they see as representations of ``truth,'' so Watterson recommends that scholarly/educational interactive media use paradata to document the ambiguous and interpretive intellectual processes that go into digital projects\cite{Watterson_DigitDwell_2015}. 

How much does this matter for a game intended for children or teens? Paradata is imperative for interactive models, games, and simulations shared among scholars because it is the only means they have to evaluate one another’s digital scholarship. I would argue that some form of paradata, however, is still necessary for children to guide them in discerning what is educationally ``factual'' and what is pure story. Institutions of learning such as museums/aquariums carry authoritative weight. Thus, communicating empirical ``truthiness'' versus playful speculation in those environments seems especially important. How to do this most effectively in an age appropriate way, and in a manner that enhances rather than breaks one's sense of immersion, is an additional area of study for later in the game development.  

\section{Process of Researching Game Technologies}

3D design in ``Allies of Atargatis'' is an essential component of worldbuilding. Given that this game also has an educational component, we strive to represent Rhodes’s architecture, marine species, and archaeological artifacts with as much veracity as is feasible. However, given our youth target audience, the art style is intentionally stylized rather than photorealistic. To achieve this balance of accuracy and stylization, we marry traditional 3D modeling in Blender with 3D Gaussian Splatting (3DGS) generated through the world model Marble. Experimenting with 3DGS allows us to generate a vast, walkable underwater world quickly, making it a suitable emerging technique for our prototyping stage. The rich photographic and video data that the Director gathered during fieldwork in Rhodes provided ample image prompts for Marble to generate the underwater world that serves as the game's backdrop.

The efficiency of 3DGS frees us to focus on crafting interactive game objects using traditional 3D modeling because the latter allows for manual fine-tuning of details. This precision is necessary to simulate architectural forms, and, even more critically, to preserve the biological features of interactive Eastern Mediterranean sea animals such as the octopus vulgaris, common eagle rays, loggerhead sea turtles, and fish species.

We use the Unity game engine to build the VR game for standalone VR devices. This choice offers the flexibility to showcase ``Allies'' at  aquariums without tethering to a computer. Because the game's rich interactivity requires precise control, we are implementing most interactions through the headset's handheld controllers. The controllers afford haptic feedback to provide cues for actions such as grasping pearls from oysters and petting the eagle rays. To deepen a sense of presence and immersion, we rendered a pair of 3D hands in place of the controller models. The illusion of seeing one's own hands reinforces the player's sense of narratological metalepsis that Ryan discusses, e.g. the phenomenon of crossing the boundary between the non-diegetic and diegetic environments to ``enter'' the story. 

\section{Results to Date, Next Steps}

Summer 2026 was devoted to realizing Phase 1 of an ``Allies of Atargatis'' playable prototype, including: building the marine environment; modeling the underwater sea spa and treasure chests; modeling and engineering oysters and their pearls, common eagle rays, and an octopus vulgaris (the player’s partner in reducing ocean plastic). Phase 2, to be realized in Summer 2027, will include: modeling Rhodes Harbor; modeling and engineering Atargatis, sea turtles to liberates from fishing nets, jellyfish, dolphins and oil companies’ sonically harmful air ray guns, bioluminescent salp chains to assist with carbon sequestration, vending amphorae, and lionfish.

Late in Summer 2027, we hope to iteratively playtest the prototype with youth at the Rhodes Aquarium, where we will eventually install the game. Playtesting will include a mixed method study of qualitative questionnaires and biometric data to evaluate the efficacy of calming effects via the game's environment design and breath interface. Eventually, we would like to see this game in other Eastern Mediterranean aquariums, schools, and children’s hospitals as they are rebuilt and electricity is restored.


\begin{credits}
\subsubsection{\ackname} Northeastern University in part funded research travel to Rhodes, Greece for Casondra Sobieralski, who is the Director, Content Researcher, Lead Interaction Designer, and Lead Narrative Designer of ``Allies of Atargatis.'' Northeastern University also funded Hector Fan, ``Allies of Atargatis'' Game Engineer and 3D modeler, to work as a 2026 summer PhD student researcher for the project. Special thanks to Rhodes Hydrobiological Station researchers and staff, especially Vivian Louizidou, Gerasimos Kondylatos, and Georgios-Angelos Hatiris. Also thank you to Dr. Beverly Goodman Tchernov, Marine Geophysical Archaeologist at Charney School of Marine Sciences in the Strauss Department of Marine Geosciences at University of Haifa, and Dr. Shawn Safford, Pediatric Surgeon at University of Pittsburgh Medical Center, for their support, encouragement, and subject matter counsel. 

\subsubsection{\discintname}
The authors have no competing interests to declare that are relevant to the content of this article.
\end{credits}

\bibliographystyle{splncs04}
\bibliography{References}
\end{document}